\documentclass{article}
\usepackage[T1]{fontenc}
\usepackage[latin9]{inputenc}
\usepackage{geometry}
\usepackage[english]{babel}
\usepackage{float}
\usepackage{amsmath}
\usepackage{amsthm}
\usepackage{amssymb}
\usepackage{graphicx}
\usepackage[authoryear]{natbib}
\usepackage[unicode=true,pdfusetitle,
 bookmarks=true,bookmarksnumbered=false,bookmarksopen=false,
 breaklinks=false,pdfborder={0 0 0},pdfborderstyle={},backref=false,colorlinks=false]
 {hyperref}

\makeatletter

\providecommand{\tabularnewline}{\\}

\newenvironment{lyxcode}
	{\par\begin{list}{}{
		\setlength{\rightmargin}{\leftmargin}
		\setlength{\listparindent}{0pt}
		\raggedright
		\setlength{\itemsep}{0pt}
		\setlength{\parsep}{0pt}
		\normalfont\ttfamily}%
	 \item[]}
	{\end{list}}
\theoremstyle{plain}
\newtheorem{thm}{\protect\theoremname}
\theoremstyle{definition}
\newtheorem{example}[thm]{\protect\examplename}

\usepackage[titles]{tocloft}

\usepackage{url}

\usepackage{amsmath}

\makeatother

\providecommand{\examplename}{Example}
\providecommand{\theoremname}{Theorem}

\begin{document}
\title{Profit and loss attribution: An empirical study}
\author{Solveig Flaig\thanks{Deutsche Rückversicherung AG, Kapitalanlage / Market risk management,
Hansaallee 177, 40549 Düsseldorf, Germany and Carl von Ossietzky Universität,
Institut für Mathematik, 26111 Oldenburg, Germany. E-Mail: solveig.flaig@deutscherueck.de.}, Gero Junike\thanks{Corresponding author. Carl von Ossietzky Universität, Institut für
Mathematik, 26111 Oldenburg, Germany. E-Mail: gero.junike@uol.de.}}
\maketitle
\begin{abstract}
The profit and loss (p\&l) attrition for each business year into different
risk or risk factors (e.g., interest rates, credit spreads, foreign
exchange rate etc.) is a regulatory requirement, e.g., under Solvency
2. Three different decomposition principles are prevalent: one-at-a-time
(OAT), sequential updating (SU) and average sequential updating (ASU)
decompositions. In this research, using financial market data from
2003 to 2022, we demonstrate that the OAT decomposition can generate
significant unexplained p\&l and that the SU decompositions depends
significantly on the order or labeling of the risk factors. On the
basis of an investment in a foreign stock, we further explain that
the SU decomposition is not able to identify all relevant risk factors.
This potentially effects the hedging strategy of the portfolio manager.
In conclusion, we suggest to use the ASU decomposition in practice.
\end{abstract}
\begin{lyxcode}
\end{lyxcode}

\section{Introduction}

The analysis of the profits and losses (p\&l) between two reporting
dates is a common task in risk management. The \emph{risk factors},
must be identified and their contribution to the p\&l has to be quantified.
The current regulation for insurance companies in Europe, Solvency
2, requires that when an internal model is used, a \emph{profit and
loss attribution} or \emph{change analysis} must be performed in sufficient
detail according to the risk categorization selected in the internal
model, see Article 240 in Solvency 2. Typically, for the market risk
module, the investment p\&l is allocated to the Solvency Capital Requirement
(SCR) sub-modules: interest rate, credit spread, equities, real estate
and foreign exchange. There are at least two basic requirements: a
p\&l attribution should be able to fully explain the p\&l, i.e., there
should not be any unexplained p\&l and a change analysis should identify
\emph{all} relevant risk factors.

The p\&l attribution can also have an economic impact when trading
decisions, e.g., the amount of the foreign exchange hedge, are based
on this analysis. 

In practice, there are many well-known decompositions that can be
used to identify the contribution of different risk factors: There
is the \emph{sequential updating} (SU) decomposition, which is also
known as \emph{waterfall}, see \citet{cadoni2014internal}. The SU
decomposition depends on the order or labeling of the risk factors:
if there are $d$ risk factors, there are $d!$ SU decompositions. 

The average of the $d!$ possible SU decompositions is called the
\emph{average sequential updating }(ASU) decomposition. The ASU decomposition
is also known as the \emph{Shapley value} or \emph{Shapley-Shubik}
decomposition and has many desirable properties, see \citet{friedman1999three}
and references therein. 

The \emph{one-at-a-time} (OAT) decomposition is also known as \emph{bump
and reset}, see \citet{cadoni2014internal}. In general, the OAT decomposition
is not able to fully explain the p\&l. 

All decompositions are linear; therefore, to decompose a portfolio
it is possible to decompose the instruments of the portfolio individually.
\citet{frei2020new}, \citet{jetses2022general} and \citet{christiansen2022decomposition}
applied the OAT, SU and ASU decompositions recursively to multiple
sub-intervals. In practice, one would apply a decomposition on annual,
quarterly, monthly or more granular sub-intervals.

For example, a risk manager working in an insurance company has to
make many decisions to implement a change analysis: which of the $d!+2$
decompositions (OAT, $d!$ SU or ASU) is the most appropriate? How
to choose the size of the sub-intervals?

In Section \ref{sec:bond}, we apply these decompositions to a typical
investment of an insurance company; a corporate bond. We show empirically
that the unexplained p\&l of the OAT decomposition is significant
and that the p\&l attribution differs significantly for the different
SU decompositions. A priori, it is unclear, which of the $d!$ SU
decomposition works best. 

Further, in Section \ref{sec:stock}, we provide an example of a partially
hedged portfolio with significant foreign exchange exposure, but the
OAT and some SU decompositions assign zero p\&l to the risk factor
describing the foreign exchange rate. This potentially effects the
hedging strategy of the portfolio manager: a naive SU or OAT decomposition
may lead to wrong trading and hedging decisions. The ASU decomposition
does not have this drawback. Therefore, both the OAT and the SU decompositions
are unsuitable for performing a change analysis.

From a theoretical point of view, the ASU decomposition is preferred
because it is able to fully explain the p\&l and does not depend on
the order or labeling of the risk factors. Moreover, it seems reasonable
to keep the sub-intervals as small as possible in order to take into
account the whole paths of the risk factors, see \citet{mai2023performance},
and to prevent inconsistencies when using conflicting sub-intervals
for different purposes. However, the available computational power
limits the number of the sub-intervals. 

Our empirical experiments based on a corporate bond indicate that
the ASU decomposition changes significantly using annual versus daily
data. However, the difference between the ASU decomposition based
on monthly and daily data is less than $0.2\%$ across twenty business
years and all risk factors.

In conclusion, we suggest to use the ASU decomposition and monthly
or more granular sub-intervals to obtain a change analysis of a (corporate)
bond portfolio. 

\section{\label{sec:bond}Change analysis of a bond}

Let us look at the evolution of the risk factors relevant for pricing
a USD A-rated corporate bond with a maturity of ten years. We consider
a European investor, so the p\&l has to be converted into EUR and
is expressed in percentage points of the nominal in EUR. We take into
account the time period $2002/12/31-2022/12/31$, i.e., twenty business
years. The most granular available data is daily data. 

The bond is modeled as a constant maturity\emph{ }bond, i.e., at \emph{every}
day the maturity of the bond is assumed to be $\mathcal{T}=10$ years.
This is the usual assumption if the insurance portfolio is grouped
into benchmark portfolios that are represented by a bond with an average
rating and duration. The risk factors are: interest rates (IR), credit
spread (CS) and foreign exchange rate (FX). 

The following Bloomberg tickers are used: USSW10 Index, C40410Y Index
and USDEUR Curncy, which describe the US 10-year interest rates, the
US corporate credit spreads for an A-rated class and a maturity of
ten years and the USDEUR exchange rate, respectively. The price of
the bond at time $t\geq0$ with maturity $\mathcal{T}$ at time $t$
is given by
\begin{equation}
P(r(t),s(t),x(t))=\frac{x(t)}{(1+r(t)+s(t))^{\mathcal{T}}},\label{eq:bond}
\end{equation}
where $r(t)$ and $s(t)$ denote a realization of IR and CS at time
$t$ for time horizon $\mathcal{T}$. The term $x(t)$ denotes a realization
of the foreign exchange rate between the foreign currency and the
domestic currency at time $t$. In this section, we consider only
a bond. Other instruments could be treated similarly by defining $P$
appropriately. 

Under Solvency 2, an insurance company has to explain the p\&l
\[
\Delta P=P(r(1),s(1),x(1))-P(r(0),s(0),x(0))
\]
for every business year, i.e., between two reporting dates $0$ and
$1$.

The OAT decomposition decomposes the p\&l $\Delta P$ by

\begin{align}
\Delta P= & \left\{ P(r(0),s(1),x(0))-P(r(0),s(0),x(0))\right\} \nonumber \\
 & +\left\{ P(r(1),s(0),x(0))-P(r(0),s(0),x(0))\right\} \nonumber \\
 & +\left\{ P(r(0),s(0),x(1))-P(r(0),s(0),x(0))\right\} +R,\label{eq:OAT3d}
\end{align}
where $R\in\mathbb{R}$ is the \emph{unexplained }p\&l\emph{. }The
first, second and third term in (\ref{eq:OAT3d}) corresponds to the
contribution of CS, IR and FX to the p\&l, respectively. That is,
to obtain the contribution of a risk factor, fix all other risk factors
at the origin, i.e., at time $0$ and allow only the risk factor of
interest to move from time $0$ to time $1$.

The SU decomposition is similarly defined but updates the risk factors
sequentially: after updating one risk factor, it is not reset to time
$0$ but kept at time $1$, see \citet{cadoni2014internal}. If there
are three risk factors, there are $3!=6$ SU decompositions. For example,
the three risk factors IR, CS and FX can be ordered according to Table
\ref{tab:Definition-of-order}. For example, the update order (CS,
IR, FX) results in to the following SU decomposition:
\begin{align}
\Delta P= & \left\{ P(r(0),s(1),x(0))-P(r(0),s(0),x(0))\right\} \nonumber \\
 & +\left\{ P(r(1),s(1),x(0))-P(r(0),s(1),x(0))\right\} \nonumber \\
 & +\left\{ P(r(1),s(1),x(1))-P(r(1),s(1),x(0))\right\} .\label{eq:SU3d}
\end{align}
The first, second and third term in Equation (\ref{eq:SU3d}) is interpreted
as the contributions of the risk factors CS, IR and FX, respectively.
The SU decomposition fully explains the p\&l, i.e., there is no unexplained
p\&l, because the right-hand side of Equation (\ref{eq:SU3d}) is
actually a telescoping series. The ASU decomposition is defined as
the average of all possible SU decompositions. 

One may also divide the business year into $m$ sub-intervals and
apply a static decomposition recursively along the sub-intervals to
define a decomposition in a multi-period setting, see \citet{frei2020new},
\citet{jetses2022general} and \citet{christiansen2022decomposition}.
For example, these sub-intervals can be chosen annual, quarterly,
monthly, weekly or daily. 

\begin{table}[H]
\begin{centering}
\begin{tabular}{|c|c|c|c|c|c|c|}
\hline 
update order & 1 & 2 & 3 & 4 & 5 & 6\tabularnewline
\hline 
\hline 
 & CS IR FX & IR CS FX & IR FX CS & CS FX IR & FX IR CS & FX CS IR\tabularnewline
\hline 
\end{tabular}
\par\end{centering}
\caption{\label{tab:Definition-of-order}Definition of the six possible update
orders of the risk factors IR, CS and FX to obtain six SU decompositions.}
\end{table}

In the business year 2020 (2003), the OAT decomposition has an unexplained
p\&l of $-2.4\%$ $(-0.8\%)$ using quarterly (daily) sub-intervals.
The absolute average unexplained p\&l over all twenty business years
are $0.4\%$ for annual and quarterly sub-intervals and $0.3\%$ for
monthly and daily sub-intervals. No matter which sub-intervals we
choose, the unexplained p\&l can be significant.

\begin{figure}[h]
\begin{centering}
\includegraphics[width=14cm]{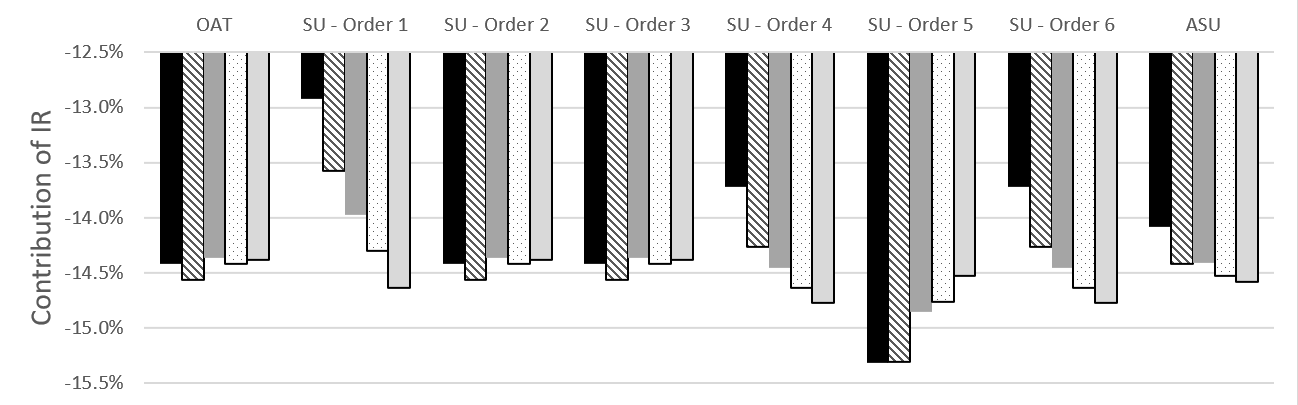}
\par\end{centering}
\begin{centering}
\includegraphics[width=14cm]{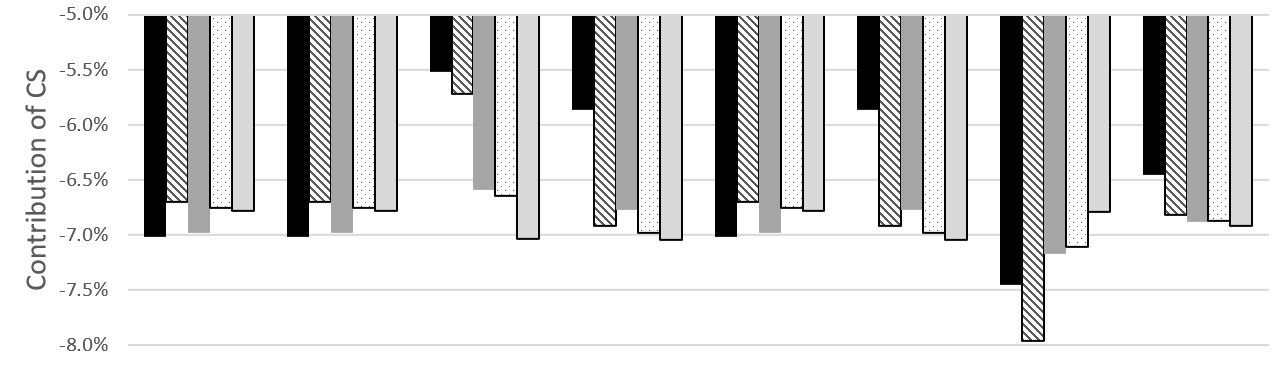}
\par\end{centering}
\begin{centering}
\includegraphics[width=14cm]{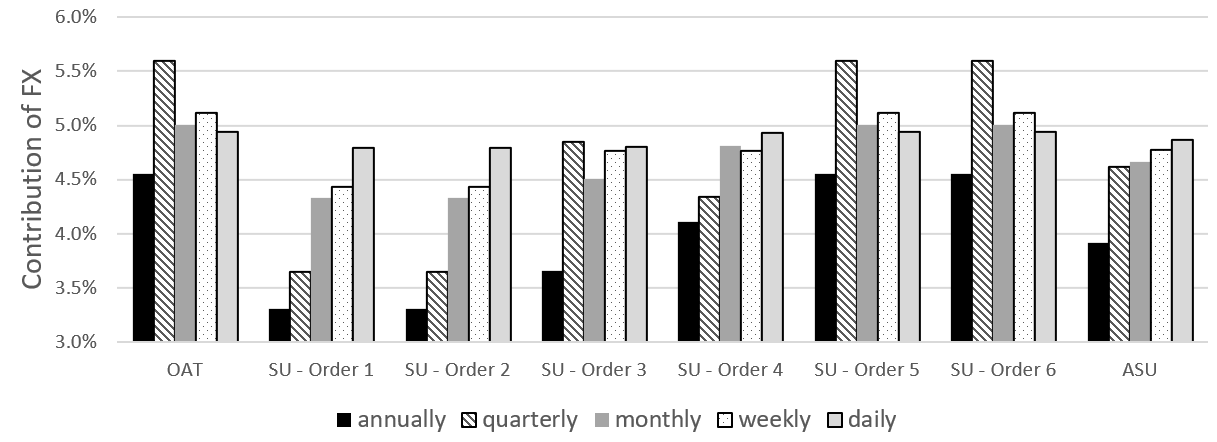}
\par\end{centering}
\caption{\label{fig:OAT_ISU_AISU_2003}OAT, SU and ASU decompositions for business
year 2022 of a US corporate bond.}
\end{figure}

Figure \ref{fig:OAT_ISU_AISU_2003} shows the six SU decompositions
for the update orders defined in Table \ref{tab:Definition-of-order}
and the business year 2022. The OAT decomposition is shown on the
left. The ASU decomposition is shown on the right. It can be seen
that the SU decomposition clearly depends on the updating order of
the risk factors, a pattern that can also be observed in other years.
The range (i.e., the difference between the largest and smallest value)
for the risk factor FX (IR) in 2022 (2003) across the six SU decompositions
using quarterly (daily) sub-intervals is $1.9\%$ $(0.9\%)$. 

On average over all twenty business years, the range across the six
possible permutations of the FX, IR, CS risk factors for monthly sub-intervals
are $0.4\%$, $0.3\%$ and $0.2\%$, respectively. No matter which
sub-intervals we choose, the SU decomposition may depend significantly
on the order or labeling of the risk factors.

The ASU decomposition is defined by the average of the six SU decompositions.
The ASU decomposition is independent of the order of the risk factors
and is able to fully explain the p\&l. Like the SU or the OAT decomposition,
the ASU decomposition depends on the size of the sub-intervals. In
2022, the range across annual, quarterly, monthly, weekly and daily
sub-intervals of the ASU decomposition for the contribution of the
risk factor FX is $1.0\%$. 

While it seems economically reasonable to use daily sub-intervals
to obtain the ASU decomposition in order to take into account the
whole paths of the risk factors, for the corporate bond we considered
in this section, it may be sufficient to use monthly sub-intervals
since the difference of the ASU decomposition based on daily versus
monthly sub-intervals is less than $0.2\%$ across all twenty business
years and the three risk factors FX, IR and CS.

\section{\label{sec:stock}SU decomposition of a US stock}

In this section, we provide an example to illustrate that a SU decomposition
may assign zero p\&l to the FX risk factor even though the portfolio
has significant FX exposure. In particular, we consider an investment
by a European investor in the US stock market such that the FX risk
is hedged by an FX-future. The FX-hedge is not perfect; some FX-risk
remains. However, there is a SU decomposition which indicates that
the p\&l is only driven by the movements of the stock market. We will
also work with real market data.

Consider a static setting and a partially hedged portfolio: Suppose
there are two risk factors, say $X$ and $Y$, and some portfolio
$P$ such that $X(t),Y(t)\in\mathbb{R}$ for $t\in\{0,1\}$ and
\begin{align}
P(X(1),Y(1)) & \neq P(X(0),Y(1)),\label{eq:cond}\\
P(X(1),Y(0)) & =P(X(0),Y(0)).\label{eq:cond2}
\end{align}
Equation (\ref{eq:cond}) says that the p\&l of $P$ depends on $X(1)$.
Equation (\ref{eq:cond2}) says that the portfolio is partially hedged
with respect to $X$: if only $X$ changes and $Y$ remains constant,
the value of the portfolio does not change.

We apply the decompositions OAT, SU and ASU to obtain the contribution
of the risk factor $X$ to a portfolio that satisfy Equations (\ref{eq:cond},\ref{eq:cond2}):
Equation (\ref{eq:OATSU1}) is the contribution of $X$ to the p\&l
according to the OAT and the SU decomposition updating the risk factor
$X$ first. Equation (\ref{eq:SU2}) is the contribution of $X$ according
to the SU decomposition updating the risk factor $Y$ first. Equation
(\ref{eq:ASU}) is the contribution of $X$ according to the ASU decomposition:
\begin{align}
P(X(1),Y(0))-P(X(0),Y(0)) & =0,\label{eq:OATSU1}\\
P(X(1),Y(1))-P(X(0),Y(1)) & \neq0,\label{eq:SU2}\\
\frac{1}{2}\left(P(X(1),Y(0))-P(X(0),Y(0))+P(X(1),Y(1))-P(X(0),Y(1))\right) & \neq0.\label{eq:ASU}
\end{align}

\begin{example}
\label{exa:SP500}We consider a portfolio $P$ in EUR consisting in
a long position in the S\&P 500, $Y$ for short, and a short position
in $Y(0)$ FX-forwards with strike $X(0)$, where $X$ models the
USDEUR exchange rate, i.e.,
\begin{equation}
P(X(t),Y(t))=X(t)Y(t)+Y(0)(X(0)-X(t)),\quad t\in\{0,1\}.\label{eq:port}
\end{equation}
$P$ satisfies Equations (\ref{eq:cond},\ref{eq:cond2}). The S\&P
500 changed in the business year 2003 from $Y(0)=880$ to $Y(1)=1110$
points. The USDEUR exchange rate changed from $X(0)=0.95$ to $X(1)=0.79$.
The portfolio in Equation (\ref{eq:port}) changed from $P(0)=836.0$
EUR to $P(1)=1017.7$ EUR, i.e., by $22\%$, while the S\&P 500 increased
by $26\%$. Therefore, the movements of $X$ must also have a significant
impact on the p\&l of $P$. However, if we apply the SU decomposition,
which updates $X$ first, or the OAT decomposition, the contribution
of $X$ is defined as zero, see Equation (\ref{eq:OATSU1}). According
to the ASU decomposition, the contribution of $X$ is given by $20.45$
EUR and the contribution of $Y$ is given by $161.25$ EUR, see Equation
(\ref{eq:ASU}). 
\end{example}

In summary, the OAT decomposition and some SU decompositions may assign
a zero contribution to the risk factor $X$, even though $X$ may
have significant impact on the p\&l. The ASU decomposition assigns
a non-zero contribution to $X$. This is potentially relevant to an
insurance company because the decomposition of the p\&l usually has
some impact on the portfolio manager's trading and hedging strategy.

\subsection*{Acknowledgments and funding}

We thank Marcus Christiansen for very fruitful discussions that helped
to improve this letter. S. Flaig would like to thank Deutsche Rückversicherung
AG for the funding for this research. Opinions, errors and omissions
are solely those of the authors and do reflect on Deutsche Rückversicherung
AG or its affiliates.

\subsection*{Disclosure statement. }

The authors report no competing interests to declare. 
\begin{lyxcode}
\bibliographystyle{plainnat}
\phantomsection\addcontentsline{toc}{section}{\refname}\bibliography{biblio}

\begin{thebibliography}{6}
\providecommand{\natexlab}[1]{#1}
\providecommand{\url}[1]{\texttt{#1}}
\expandafter\ifx\csname urlstyle\endcsname\relax
  \providecommand{\doi}[1]{doi: #1}\else
  \providecommand{\doi}{doi: \begingroup \urlstyle{rm}\Url}\fi

\bibitem[Cadoni(2014)]{cadoni2014internal}
Paolo Cadoni.
\newblock \emph{Internal models and Solvency II}.
\newblock Risk Books, London, 2014.

\bibitem[Christiansen(2022)]{christiansen2022decomposition}
Marcus~C Christiansen.
\newblock {On the decomposition of an insurer's profits and losses}.
\newblock \emph{Scandinavian Actuarial Journal}, pages 1--20, 2022.

\bibitem[Frei(2020)]{frei2020new}
Christoph Frei.
\newblock A new approach to risk attribution and its application in credit risk
  analysis.
\newblock \emph{Risks}, 8\penalty0 (2):\penalty0 65, 2020.

\bibitem[Friedman and Moulin(1999)]{friedman1999three}
Eric Friedman and Herve Moulin.
\newblock {Three methods to share joint costs or surplus}.
\newblock \emph{Journal of economic Theory}, 87\penalty0 (2):\penalty0
  275--312, 1999.

\bibitem[Jetses and Christiansen(2022)]{jetses2022general}
Julian Jetses and Marcus~C Christiansen.
\newblock A general surplus decomposition principle in life insurance.
\newblock \emph{Scandinavian Actuarial Journal}, 2022\penalty0 (10):\penalty0
  901--925, 2022.

\bibitem[Mai(2023)]{mai2023performance}
Jan-Frederik Mai.
\newblock Performance attribution with respect to interest rates, fx, carry,
  and residual market risks.
\newblock \emph{arXiv preprint arXiv:2302.01010}, 2023.

\end{thebibliography}

\end{lyxcode}

\end{document}